\documentclass[11pt]{article}
\usepackage{geometry}               
\usepackage{graphicx}
\usepackage{epstopdf}
\def\be{\begin{equation}}
\def\ba{\begin{array}}
\def\bfg{\begin{figure}}
\def\ef{\end{figure}}
\def\bea{\begin{eqnarray*}}		
\def\ee{\end{equation}}
\def\ea{\end{array}}
\def\eea{\end{eqnarray*}}

\begin{document} 
\title{Monte Carlo and kinetic Monte Carlo methods -- a tutorial}
\author{Peter Kratzer \\
Fachbereich Physik and Center for Nanointegration (CeNIDE), \\
Universit\"at Duisburg-Essen, Lotharstr. 1, 47048 Duisburg, Germany}
\date{}

\maketitle

\begin{abstract}
This article reviews the basic computational techniques for carrying out multi-scale simulations using statistical methods, with the focus on simulations of epitaxial growth. 
First, the statistical-physics background behind Monte Carlo simulations is briefly described. The kinetic Monte Carlo (kMC) method is introduced as an extension of the more wide-spread thermodynamic  Monte Carlo methods, and algorithms for kMC simulations, including parallel ones, are discussed in some detail. The step from the atomistic picture to the more coarse-grained description of Monte Carlo simulations is exemplified for the case of surface diffusion. Here, the aim is the derivation of rate constants from knowledge about the underlying atomic processes. Both the simple approach of Transition State Theory, as well as more recent approaches using accelerated molecular dynamics are reviewed. Finally, I address the point that simplifications  often need to be introduced in practical Monte Carlo simulations in order to reduce the complexity of 'real' atomic processes. Different 'flavors' of kMC simulations and the potential  pitfalls related to the reduction of complexity are presented in the context of simulations of epitaxial growth.   
\end{abstract}

\section{Introduction}

In Computational Materials Science we have learned a lot from molecular dynamics (MD) simulations that allows us to follow the dynamics of molecular processes in great detail. In particular, the combination of MD simulations with density functional theory (DFT) calculations of the electronic structure, as pioneered more than thirty years ago by the work of R. Car and M. Parrinello\cite{CaPa85}, has brought us a great step further:  Since DFT enables us to describe a wide class of chemical bonds with good accuracy, it has become possible to model the microscopic dynamics behind many technological important  processes in materials processing or chemistry. 
It is important to realize that the knowledge we gain by interpreting the outcome of a simulation can be only as reliable as the theory at its  basis that solves the quantum-mechanical problem of the system of electrons and nuclei for us. Hence any simulation that aims at predictive power should start from the sub-atomic scale of the ele
ctronic many-particle problem. 
However, for many questions of scientific or technological relevance, the phenomena of interest take place on much larger length and time scales. 
Moreover, temperature may play a crucial role, for example in phase transitions. Problems of this type have been handled by Statistical Mechanics, and special techniques such as Monte Carlo methods have been developed to be able to tackle with complex many-particle systems\cite{PaBa99,LaBi00}. However, in the last decade it has been realized that also for those problems that require statistics for a proper treatment, a 'solid basis' is indispensable, i.e. an understanding of the underlying molecular processes, as provided by DFT or quantum-chemical methods.
This has raised interest in techniques to combine Monte Carlo methods with a realistic first-principles description of processes in condensed matter.\cite{KrSc:01a}
 
I'd like to illustrate these general remarks with examples from my own field of research, the theory of epitaxial growth. The term epitaxy means that the crystalline substrate imposes its structure onto some deposited material, which  may form a smooth film or many small islands, depending on growth conditions. 
Clearly, modeling the deposition requires a sample
area of at least mesoscopic size, say 1~$\mu$m$^2$, involving tens of 
thousands of atoms. The time scale one would like to cover by the 
simulation should be of the same order as the actual time used to
deposit one atomic layer, i.e. of the order of seconds.  
\begin{figure}[t]
\centerline{   
\includegraphics[width=7.0cm]{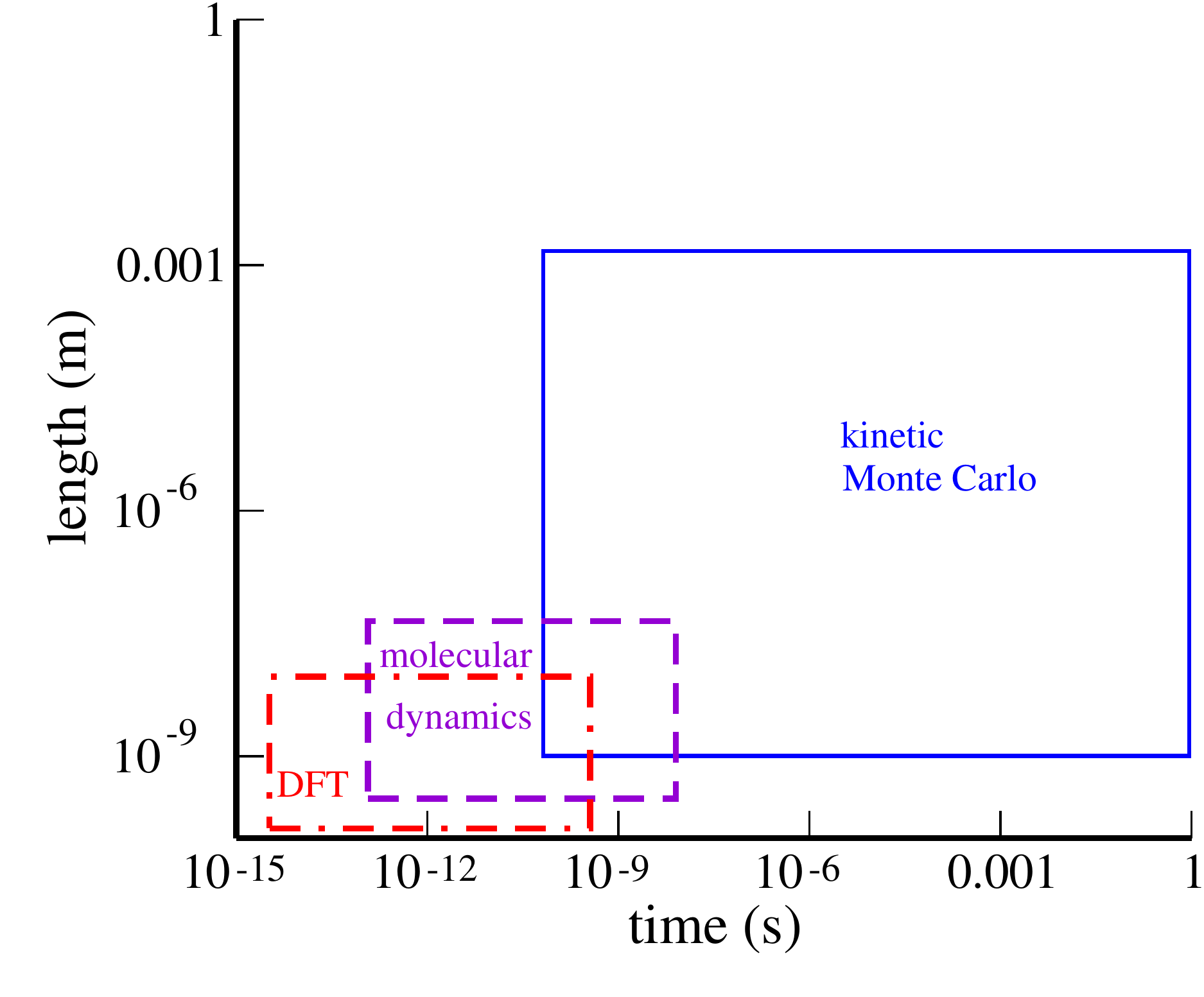}
}
\caption{Molecular modeling on the basis of first-principles
electronic structure calculations requires to cover the length and time
scales from the electronic to the mesoscopic or even macroscopic regime.
On the electronic level, density-functional theory (DFT) is frequently employed. 
Molecular dynamics (MD) simulations can be carried out either in combination with DFT, or by using classical forces, which allow one to extend the simulations to bigger length and time scales. The kinetic Monte Carlo method may reach out to very large scales (much depending on the rate constants of the processes relevant to a specific problem), while being able to use input from DFT or MD simulations. }
\label{scales}
\end{figure}
However, the microscopic, atomistic processes that govern the physics
and chemistry of deposition, adsorption, and diffusion operate in the 
length and time domains of 0.1 to 1 nm, and femto- to pico-seconds.
Hence incorporating information about atomic processes into modeling of film growth poses the challenge to cover huge length and time 
scales: from $10^{-10}$m to $10^{-6}$m and from $10^{-15}$s to $10^0$ s
(cf. Fig.~\ref{scales}).
While smaller length scales, comprising a few hundred atoms, are
typically sufficient to gain insight, e.g. about the atomic 
structure of a step on a surface and its role for chemical reactions
and atom diffusion, the gap between the atomic and the practically 
relevant {\em time scales} and the crucial role of Statistical
Mechanics constitute major obstacles for reliable molecular 
modeling. 

An additional challenge arises due to the complexity of the phenomena to be investigated: 
One of the fascinating features of epitaxy is the {\em  interplay} of various atomic processes. For example, atoms deposited on an island may be able to overcome the island edge ('step down to the substrate') for specific edge orientations. Thus, while the island changes its shape (and thus the structure of its edges) during growth, this will enable (or disable) material transport between the island top and the substrate, resulting in a transition from two-dimensional to three-dimensional island growth (or vice versa). 
The possibility that processes may 'trigger' other processes during the evolution of structures can hardly be foreseen or incorporated {\em a priori} in analytical modeling, but calls for computer simulations using statistical methods. 

In epitaxial growth, lattice methods exploiting the two-dimensional periodicity of the substrate lattice are often -- but not always -- appropriate. Also in other fields of Solid State Physics, mathematical models defined on lattices have been used for a long time.  A well-known example is the Ising model in the study of magnetism. It describes the interaction between magnetic moments (spins) sitting on a lattice that can take on two states only ('up' or 'down', represented by variables $s_i= \pm 1$).
The Hamiltonian of the Ising model is given by 
\be
H(s) = -J_q \sum_i \sum_{j \in n(i)} s_i s_j - \mu_B B \sum_i s_i \,
\ee
where 
$n(i)$ denotes the set of spins
interacting with spin $i$, $J_q$ is the strength of the interaction between
spins, $q$ is the number of interacting neighbors ($q J_q = \mathrm{const} = k_BT_c$, where the last equality is valid in the mean-field approximation), and $B$ is an external magnetic field .

In surface physics and epitaxy, a mathematically equivalent model is used under the name 'lattice Hamiltonian'. It describes fixed sites on a lattice that can be either empty or occupied by a particle (e.g., a deposited atom). The interactions between these particles are assumed to have finite range. 
The lattice-gas interpretation of the Ising model is obtained by the
transformation $s_i = 2c_i-1, \quad c_i = 0, 1$,
\be 
H = -4J_q \sum_i \sum_{j \in n(i)} c_i c_j + 2(q J_q - \mu_B B) \sum_i c_i - N(q J_q - \mu_B B) \, .
\ee

For studies of epitaxy, one wishes to describe not only monolayers of atoms, but films that are several atomic layers thick. These layers may not always be complete, and islands and steps may occur on the growing surface. For a close-packed crystal structure, the atoms at a step are chemically less coordinated, i.e., they have fewer partners to form a chemical bond than atoms sitting in flat parts of the surface (= terraces), or atoms in the bulk. Hence, it costs additional energy to create steps, or kinks in the steps. Inspired by these considerations, one can define the so-called solid-on-solid (SOS) model, in which each lattice site is associated with an integer variable, the local surface height $h_i$. In the 
SOS model, an energy 'penalty' must be paid whenever two neighbouring lattice sites differ in surface height,
\be 
H = K_q \sum_i \sum_{j \in n(i)} |h_i - h_j| \, .
\ee
This reflects the energetic cost of creating steps and kinks. 
The SOS model allows for a somewhat idealized, but still useful description of the morphology of a growing surface, in which the surface can be described mathematically by a single-valued function $h$ defined on a lattice, i.e., no voids or overhangs in the deposited material are allowed. 
In the following, we will sometimes refer to one of these three models to illustrate certain features of Monte Carlo simulations. 
More details about these and other models of epitaxial growth can be found in books emphasizing the statistical-mechanics aspects of epitaxy, e.g. in the textbook by Stanley and Barabasi\cite{BaSt95}.

In studies of epitaxial growth, model systems defined through a simple Hamiltonian, such as the lattice-gas or the SOS model, have a long history, and numerous phenomena could be described using kinetic Monte Carlo simulations based on these models, dating back to 
early work by G. H. Gilmer\cite{Gilm76}, later extended by D. D. Vvedensky\cite{clarke:87} and others. 
For the reader interested in the wealth of structures observed in the 
evolution of surface morphology, I recommend the book by T. Michely and J. Krug\cite{MiKr04}. 
Despite the rich physics that could be derived from simple models, research in the last decade has revealed that such models are still too narrow a basis for the processes in epitaxial growth.
Thanks to more refined experimental techniques, in particular scanning tunneling microscopy\cite{ZhLa97}, but also thanks to atomistic insights provided by DFT calculations\cite{BoSt98,OvBo99}, we have learned in the last ten years that the processes on the atomic scale are by no means simple. For example, the numerous ways how atoms may attach to an island on the substrate display a stunning  complexity. However, kinetic Monte Carlo methods are flexible enough so that the multitude of possible  atomic processes can be coded in a simulation program easily, and their macroscopic consequences can be explored.

Apart from simulations of epitaxial growth, thermodynamic as well as  kinetic Monte Carlo simulations are a valuable tool in many other areas of computational physics or chemistry. 
In polymer physics, the ability of Monte Carlo methods to bridge time and length scales makes them very attractive: For example, the scaling properties of polymer dynamics on long time scales (often described by power laws) can be investigated by Monte Carlo simulations.\cite{KoTh04} 
Another important field of applications is in surface chemistry and catalysis\cite{NeHa00,BrSn00}: Here, Monte Carlo methods come with the bargain that they allow us to study the interplay of a large number of chemical reactions more easily and reliably than the traditional method of rate equations. 
Moreover, also in this field, feeding information about the individual molecular processes, as obtained e.g. from DFT calculations, into the simulations is a modern trend pursued by a growing number of research groups\cite{ReSc06,SeSa06}.

\section{Monte Carlo methods in Statistical Physics}
The term 'Monte Carlo' (MC) is used for a wide variety of methods in theoretical physics, chemistry, and engineering where random numbers play an essential role. 
Obviously, the name alludes to the famous casino in Monte Carlo, where random numbers are generated by the croupiers (for exclusively non-scientific purposes). In the computational sciences, we generally refer to 'random' numbers  generated by a computer, so-called quasi-random numbers
\footnote{Concerning the question how a deterministic machine, such as a computer, could possibly generate 'random' numbers, the reader is referred to the numerical mathematics literature, e.g. Ref.~\cite{Pres86}.}. 

A widely known application of random numbers is the numerical evaluation of intergrals in a high-dimensional space. There, the integral is replaced by a sum over function evaluations at discrete support points. 
These support points are drawn from a random distribution in some compact $d$-dimensional support ${\cal C}$. 
If the central limit theorem of statistics is applicable, the sum converges, in the statistical sense, towards  the value of the integral.   The error decreases proportional to the inverse square root of the number of support points, independent of the number of space dimensions. Hence, Monte Carlo integration is an attractive method in particular for integration in high-dimensional spaces.

In Statistical Physics, a central task is the evaluation of the partition function of the canonical ensemble for an interacting system, described by a Hamiltonian $H$. The contribution of the kinetic energy to $H$ is simple, since it is a sum over single-particle terms. However, calculating the potential energy term  
$U(x)$ for an interacting many-particle system involves the evaluation of  a high-dimensional integral of the type 
\be
Z = \int_{\cal C} dx \, \exp\left( -\frac{U(x)}{k_{\mathrm B} T} \right) \, .
\ee
Here, $x$ stands for a high-dimensional variable specifying the system configuration (e.g., position of all particles).
Evaluating this integral by a Monte Carlo method requires special care: 
Only regions in space where the potential $U$ is small contribute strongly. Hence, using a {\it uniformly} distributed set of support points would waste a lot of computer resources. Instead, one employs a technique called {\bf importance sampling}. 
We re-write the partition function
\be
Z = \int_{\cal C} d\mu(x) 
\ee
with the Gibbs measure $d\mu(x)=\exp (-U(x)/(k_{\rm B} T) ) \, dx$. 
The expectation value for an observable is evaluated as the sum over $n$ sampling points in the limit of very dense sampling, 
\be
\langle O \rangle  = \frac{1}{Z} \int_{\cal C} O(x) \, d\mu(x) 
= \lim_{n \to \infty} \frac{ \sum_{i=1}^{n} O(x_i) \mu(x_i)}{\sum_{i=1}^{n} \mu(x_i)} \, .
\ee
When we generate the $n$ sampling points in configuration space according to their
equilibrium distribution,  
$P_{\mathrm{eq}}(x) = {1 \over Z} \exp(- U(x)/(k_{\rm B} T)) \approx \mu(x_i)/\sum_{i=1}^{n} \mu(x_i)$,
we are in position to calculate the thermodynamic average of any observable using
\be
\langle O \rangle \approx \frac{1}{n} \sum_{i=1}^{n} O(x_i) \, .
\ee
The remaining challenge is to generate the support points according to the equilibrium distribution. Instead of giving an explicit description of the equilibrium distribution, it is often easier to think of a stochastic process that tells us how to build up the list of support points for the Gibbs measure. If an algorithm can be judiciously designed in such a way as to retrieve the equilibrium distribution as its limiting distribution, knowing this algorithm (how to add support points) is as good as knowing the final outcome. This 'philosophy' is behind many applications of Monte Carlo methods, both in the realm of quantum physics (Quantum Monte Carlo) and in classical Statistical Physics. 

To be more precise, we have to introduce the notion of a {\bf Markov process}. Consider that the system is in a generalized state $x_i$ at some time $t_i$. (Here, $x_i$ could be a point in a $d$-dimensional configuration space.) A specific evolution of the system may be characterized by a probability $P_n(x_1,t_1; \ldots ; x_n,t_n)$ to visit all the points $x_i$ at times $t_i$. 
For example, $P_1(x;t)$ is just the probability of finding the system in configuration $x$ at time $t$. 
Moreover, we need to introduce conditional probabilities $p_{1| n} (x_n,t_n | x_{n-1},t_{n-1}; \ldots ; x_1,t_1)$
The significance of these quantities is the probablity of finding the system at $(x_n,t_n)$ provided that it has 
visited already all the space-time coordinates $(x_{n-1},t_{n-1}) \ldots (x_1,t_1)$.   
The characteristic feature of a Markov process is the fact that transitions depend on the {\em previous} step in the chain of events {\em only}. Hence it is sufficient to consider only {\em one} conditional probablity $p_{1 | 1}$ for transitions between subsequent points in time. The total probability can then be calculated from the preceeding ones,
\be
P_n(x_1,t_1; \ldots ; x_n,t_n) = p_{1 | 1} (x_n,t_n | x_{n-1},t_{n-1}) P_{n-1}(x_1,t_1; \ldots ; x_{n-1},t_{n-1})
\ee
In discrete time, we call such a process a Markov chain. The conditional
probabilities of Markov processes obey the {\bf Chapman-Kolmogorov} 
equation
\be
p_{1|1}(x_3, t_3 | x_1, t_1) = \int dx_2 \, p_{1|1}(x_3, t_3 | x_2, t_2) p_{1|1}(x_2, t_2 | x_1, t_1)
\ee
If the Markov process is  {\em stationary}, we can write for its two defining functions
\bea
P_1(x, t) &=& P_{\mathrm{eq}}(x) ; \\
p_{1|1}(x_2, t_2 | x_1, t_1) &=& p_t(x_2 | x_1) ; \qquad t = t_2-t_1 \, .
\eea
Here $P_{\mathrm{eq}}$ is the distribution in thermal equilibrium, and $p_t$ denotes the  transition probability within the time interval $t $ from a state $x_1$ to a state $x_2$.

Using the Chapman-Kolmogorov equation for $p_t$, we get
\be
p_{t+t_0}(x_3 | x_1) = \int dx_2\, p_{t_0}(x_3 | x_2) p_t(x_2 | x_1 ) \, .
\ee
When we consider a discrete probability space for $x_i$, the time evolution of the probability proceeds by matrix multiplication, the $p_t$ being matrices transforming one discrete state into another.
We now want to derive the differential form of the Chapman-Kolmogorov equation for stationary Markov processes. Therefore 
we consider the case of small time intervals $t_0$ and write the transition probability in the following way, 
\be
p_{t_0}(x_3 | x_2) \approx (1 - w_{\mathrm{tot}}(x_2) t_0) \delta(x_3 - x_2) + t_0 w(x_3 | x_2) + \ldots , 
\ee
up to to terms that vanish faster than linear in $t_0$. This equation defines $w(x_3 | x_2)$ as the transition rate (transition
probability per unit time) to go from $x_2$ to $x_3$. In the first term, the factor $(1 - w_{\mathrm{tot}}(x_2) t_0)$ signifies the probability to remain in state $x_2$ up to time $t_0$. That means that $w_{\rm tot}(x_2)$ is the total probability to leave the state $x_2$, defined as  
\be
w_{\mathrm{tot}}(x_2) = \int  dx_3 \, w(x_3 | x_2) .
\ee
Inserting this into the Chapman-Kolmogorov equation results in
\be
p_{t+t_0}(x_3 | x_1) = (1 - w_{\mathrm{tot}}(x_3) t_0) p_t(x_3 | x_1) +t_0
\int dx_2 \, w(x_3 | x_2) p_t(x_2 | x_1) ;
\ee
and hence we obtain 
\be 
\frac{p_{t+t_0}(x_3 | x_1) - p_t(x_3 | x_1)}{t_0} =
\int dx_2 \, w(x_3 | x_2) p_t(x_2 | x_1) - 
\int dx_2 w(x_2 | x_3) p_t(x_3 | x_1) ,
\ee
in which we have used the definition of $w_{\mathrm{tot}}$. In the limit $t_0 \to 0$ we arrive at the {\bf master equation}, that is the differential version of the Chapman-Kolmogorov equation, 
\be
\frac{\partial}{\partial t} p_t(x_3 | x_1) =
\int dx_2 \, w(x_3 | x_2) p_t(x_2 | x_1) - 
\int dx_2 w(x_2 | x_3) p_t(x_3 | x_1) \, .
\label{master_eq}
\ee
It is an integro-differential equation for the transition probabilities of
a stationary Markov process.
In the following we do not assume stationarity and choose a $P_1(x_1, t) \neq 
P_{\mathrm{eq}}(x)$, but keep the assumption of time-homogeneity of the transition probabilities, i.e., it is assumed that they only depend on time differences. Then, we can multiply this equation by $P_1(x_1,  t)$ and integrate over $x_1$  to get a master equation for the probability density itself:
\be
\frac{\partial}{\partial t} P_1(x_3,t) =
\int dx_2 \, w(x_3 | x_2) P_1(x_2,t) - 
\int dx_2 w(x_2 | x_3) P_1(x_3, t)
\ee
One way to fulfill this equation is to require {\bf detailed balance}, i.e., the
net probability flux between every pair of states in equilibrium is zero, 
\be
\frac{ w(x | x')}{ w(x' |x)} 
= \frac{P_{\mathrm{eq}}(x)}{P_{\mathrm{eq}}(x')} \, .
\label{eq:detailedB}
\ee
For thermodynamic averages in the canonical ensemble, 
$P_{\mathrm{eq}}(x)= {1\over Z} \exp (- H(x)/(k_{\rm B}T) )$, and hence 
\be
\frac{ w(x | x')}{ w(x' |x)} 
= \exp\left( - (H(x) - H(x')) / (k_{\rm B} T) \right) \, .
\ee
When we use transition probabilities in our Monte Carlo simulation
that fulfill detailed balance with the desired equilibrium distribution, 
we are sure to have
\be
P_1(x, t \to \infty) = P_{\mathrm{eq}}(x) \, .
\ee
Since the detailed balance condition can be fulfilled in many ways, the choice of transition rates is therefore not unique.
Common choices for these rates are
\begin{itemize}
\item the {\bf Metropolis rate}

$w(x' | x) = w_0(x' |x) \mathrm{min} (\left[ 1; \exp\left( - (H(x') - H(x)) /(k_{\rm B} T)  \right) \right]$
\item the {\bf Glauber rate} 

$w(x' | x) = w_0(x' |x) {1 \over 2} \left\{ 1 - \tanh\left[ \exp\left( - (H(x') - H(x) )/(k_{\rm B} T) \right) \right] \right\}$
\end{itemize}
Both choices obey the detailed balance condition. 
With either choice,  we still have the freedom to select a factor $w_0(x' |x) = w_0(x|x')$. This can be interpreted as the probability to choose a pair of states $x, \, x'$ which are connected
through the specified move.   In an Ising model simulation, each state $x$ corresponds to one particular arrangement of all spins on all the lattice sites. The states $x$ and $x'$ may, for instance, just differ in the spin orientation on one lattice site. Then, the freedom in $w_0(x'|x)$ corresponds to the freedom to select any single spin (with a probability of our choice), and then to flip it (or not to flip it) according to the prescription of the rate law.

Let's illustrate the general considerations by an example. Suppose we want to calculate the magnetization of an  Ising spin model at a given temperature. Hence we have to simulate the canonical ensemble using the {\bf Monte Carlo algorithm}. The steps are: 
\begin{itemize}
\item generate a starting configuration $s_0$,
\item select a spin, $s_i$, at random,
\item calculate the energy change upon spin reversal $\Delta H$,
\item calculate the probability $w(\uparrow, \downarrow)$ for this spin-flip to happen, using the chosen form of transition probability (Metropolis or Glauber),
\item generate a uniformly distributed random number, $0 < \rho < 1$; if $w > \rho$, flip the spin, otherwise retain the old configuration.
\end{itemize}
When the Metropolis rate law has been chosen, proposed spin flips are either accepted with probability $w$, or discarded with probability $1-w$. 
After some transient time, the system will come close to thermodynamic equilibrium. Then we can start to record time averages of some observable $O$ we are interested in, e.g., the magnetization. Due to the in-built properties of the rate law, this time average will converge, in the statistical sense, to the thermodynamic ensemble average $\langle O \rangle$  of the observable $O$. 

The prescription for the Monte Carlo method given so far applies to non-conserved observables  (e.g., the magnetization). For a conserved quantity (e.g., the concentration of particles), one uses {\bf Kawasaki dynamics}:
\begin{itemize}
\item choose a pair of (neighboring) spins
\footnote{This means $w_0(s' | s) = 1/(2dn)$, i.e. we first 
choose a spin at random, and then a neighbor on a $d$-dimensional simple cubic lattice with $n$ sites at random.}
\item exchange the spins subject to the Metropolis acceptance criterion\end{itemize}
Since this algorithm guarantees particle number conservation, it recommends itself for the lattice-gas interpretation of the Ising model.
In simulations of epitaxial growth, one may work with either conserved or  non-conserved particle number, the latter case mimicking desorption or adsorption events of particles.

\begin{figure}[htbp] 
   \centering
   \includegraphics[width=7cm]{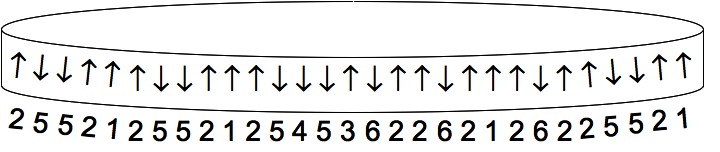} 
   \caption{Illustration of the $N$-fold way algorithm for a one-dimensional Ising chain of spins. A particular configuration for one moment in time is shown. The local environments of all spins fall in one of six classes, indicated by the numbers. Periodic boundary conditions are assumed.}
   \label{fig:BKL}
\end{figure}

\begin{table}[htdp]
\caption{Classification on spins in a 6-fold way for a periodic Ising chain. The leftmost column gives the number $n_i$ of spins in each class for the particular configuration shown in Fig.~\protect\ref{fig:BKL}. The rates can be normalised to unity by setting $w_0 = \{ [n_1 \exp(-2\mu_B/(k_{\rm B}T) ) + n_4 \exp(2\mu_B/(k_{\rm B}T) )  ]  \exp(-4J_q/(k_{\rm B} T) ) + n_2 \exp(-2\mu_B/(k_{\rm B}T) ) + n_5 \exp(2\mu_B/(k_{\rm B}T) )   + [n_3 \exp(-2\mu_B/(k_{\rm B}T) ) + n_6 \exp(2\mu_B/(k_{\rm B}T) )  ]   \exp(4J_q/(k_{\rm B} T) ) \}^{-1}$. }
\begin{center}
\begin{tabular}{|c|c|c|c|c|}
\hline
class & central & neighbors & rate & class \\
    & spin     &                     & $w_i$ &  members $n_i$ \\
\hline
1 & $\uparrow$ & $\uparrow, \uparrow$ & $w_0 \exp\bigl(-(4J_q+2\mu_B B)/(k_{\rm B} T) \bigr)$ &  4 \\
2 & $\uparrow$ & $\uparrow, \downarrow$ & $w_0 \exp\bigl(-2\mu_B B/(k_{\rm B} T) \bigr)$  & 12 \\
3 & $\uparrow$ & $\downarrow, \downarrow$ & $w_0  \exp\bigl( (4J_q-2\mu_B B)/(k_{\rm B} T) \bigr)$ & 1 \\
4 & $\downarrow$ & $\downarrow, \downarrow$ & $w_0 \exp\bigl(-(4J_q -2\mu_B B)/(k_{\rm B} T) \bigr)$ &  1 \\
5 & $\downarrow$ & $\uparrow, \downarrow$ & $w_0 \exp\bigl( 2\mu_B B/(k_{\rm B} T) \bigr)$  & 8 \\
6 & $\downarrow$ & $\uparrow, \uparrow$ & $w_0  \exp\bigl( (4J_q + 2\mu_B B)/(k_{\rm B} T) \bigr)$ & 3 \\
\hline
\end{tabular}
\end{center}
\label{tab:BKL}
\end{table}

\section{From MC to kMC: the $N$-fold way}
The step forward from the Metropolis algorithm to the algorithms used for kinetic Monte Carlo (kMC)  simulations originally resulted from a proposed speed-up of MC simulations: In the Metropolis algorithm, trial steps are sometimes discarded, in particular if the temperature is low compared to typical interaction energies. 
For this  case, Bortz, Kalos and Lebowitz suggested in 1975 the $N$-fold way algorithm\cite{BoKa75} that avoids discarded attempts.
The basic idea is the following: In an Ising model or similar models, the interaction energy, and thus the rate of spin-flipping, only depends on the nearest-neighbor configuration of each spin. Since the interaction is short-ranged, there is only a small number of local environments (here: spin triples), each with a certain rate $w_i$ for flipping the 'central' spin of the triple. 
For example, in one dimension, i.e. in an Ising chain, an 'up' spin may have both neighbors pointing 'up' as well, or both neighbors pointing 'down', or alternate neighbors, one 'up', one 'down'. An analogous classification holds if the selected (central) spin points 'down'. All local environments fall into one of these six classes, and there are six 'types' of spin flipping with six different rate constants. For a given configuration of a spin chain, one can enumerate how frequently each class of environment occurs, say, $n_i$ times, $i=1,\ldots 6$.   This is illustrated in Table~\ref{tab:BKL} for the configuration shown in Fig.~\ref{fig:BKL}.  Now the {\bf N-fold way algorithm} works like this:
\begin{enumerate}
\item first select a class $i$ with a probability given by $n_i w_i/\sum_i w_i n_i$ using a random number $\rho_1$;
\item then, select one process (i.e., one spin to be flipped) of process type $i$, choosing with equal probability among the representatives of that class, by using another random number $\rho_2$;
\item execute the process, i.e. flip the spin;
\item update the list of $n_i$ according to the new configuration.
\end{enumerate}
The algorithm cycles through this loop many times, without having to discard any trials, thereby reaching thermal equilibrium in the spin system. The prescription can easily be generalized to more dimensions; e.g., to a two-dimensional square lattice, where we have ten process types.\footnote{Each 'central' spin has four neighbors, and the number of neighbors aligned with the 'central' spin may vary between 0 and 4. Taking into account that the central spin could be up or down, we end up with ten process types.}

To go all the way from MC to kMC, what is still missing is the aspect of {\em temporal evolution}. In a MC simulation, we may count the simulation steps. However, the foundation of the method lies in  {\em equilibrium} statistical physics. Once equilibrium is reached, time has no physical meaning. Therefore no physical basis exists for identifying simulation steps with physical time steps in the conventional Monte Carlo methods. In order to address {\em kinetics}, i.e. to make a statement how fast a system reaches equilibrium, we need to go beyond that, and take into account for the role of time. 
Here, some basic remarks are in place. In order to be able to interpret the outcome of our simulations, we have to refer to some assumptions about the {\em separation of time scales}: The shortest time scale in the problem is given by the time it takes for an elementary process (e.g., a spin flip) to proceed. This time scale should be clearly separated from the time interval between two processes taking place at the same spin, or within the local environment of one spin. This second time scale is called the waiting time between two subsequent events. If the condition of time scale separation is not met, the remaining alternative is a simulation using (possibly accelerated) molecular dynamics (see Section~\ref{sec:accelerated}). If time scale separation applies, one of the basic requirements for carrying out a kMC simulation is fulfilled. The advantage is that kMC simulations can be run to simulate much longer physical time intervals at even lower computational cost than molecular dynamics simulations (see Fig.~\ref{scales}). Moreover, one can show that the waiting time, under quite general assumptions, follows a Poissonian distribution\cite{FiWe91}.  
For the Ising chain, each process type has a different waiting time $\tau_i= w_i^{-1}$ that is proportional to  some power of $\exp(J/(k_{\mathrm{B}}T))$. 
For other applications of interest, the waiting times of various process types may be vastly different. 
In epitaxial growth, for instance, the time scale between two adsorption or desorption events is usually much longer than the time scale for surface diffusion between two adjacent sites. For macromolecules in the condensed phase, vibrational motion of a molecular side group may be fast, while a rotation of the whole molecule in a densely packed environment may be very slow, due to steric hinderance. In a kinetic simulation, we would like to take all these aspects into account. We need a simulation method that allows us to {\bf bridge time scales} over several orders of magnitude.

\begin{figure}[htbp] 
   \centering
   \includegraphics[width=13cm]{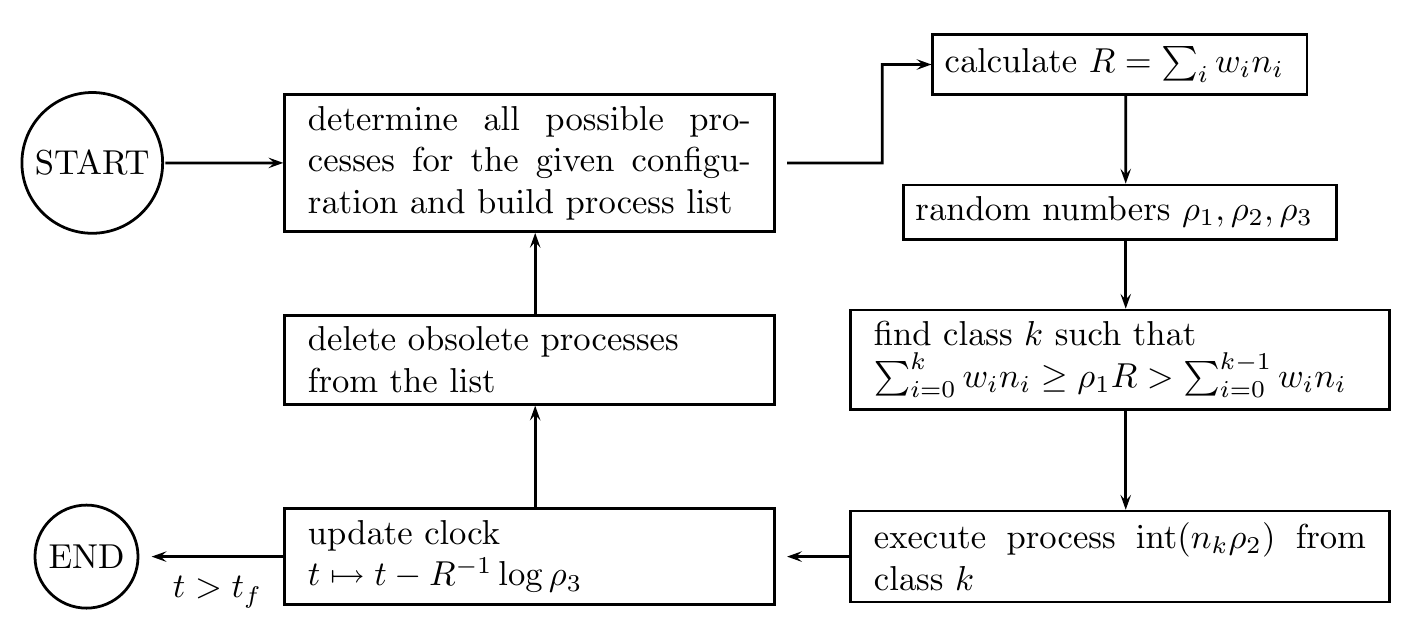} 
   \caption{Flow chart for the process-type list algorithm.}
   \label{fig:flow-chart}
\end{figure}

Following the $N$-fold way for the Ising model, it is easy to calculate the {\em total} rate $R$, i.e., the probability that some event will occur in the whole system per unit time. It is the sum of all rates of individual processes, $R= \sum_i n_i w_i$. The average waiting time between any two events occurring in the system as a whole is given by $R^{-1}$. 
This allows us to associate a time step of (on average) $\Delta t = R^{-1}$ with one step in the simulation. 
Note that the actual length of this time step may change (and in general does so) during the simulation, since the total rate of all processes accessible in a certain stage of the simulation may change. 
Therefore, this variant of the kMC method is sometimes also called the 'variable step size' method in the literature.
More realistically, the time step $\Delta t$ should not be identified with its average value, but should should be drawn from a Poissonian distribution. This is practically realised by using the expression $\Delta t = - R^{-1} \log \rho_3$ with a random number $0 < \rho_3 < 1$. 
For a two-dimensional problem (e.g., a lattice-gas Hamiltonian), the $N$-fold way algorithm is explained in the flowchart of Fig.~\ref{fig:flow-chart}.

\begin{figure}[htbp] 
   \centering
 \includegraphics[width=12cm]{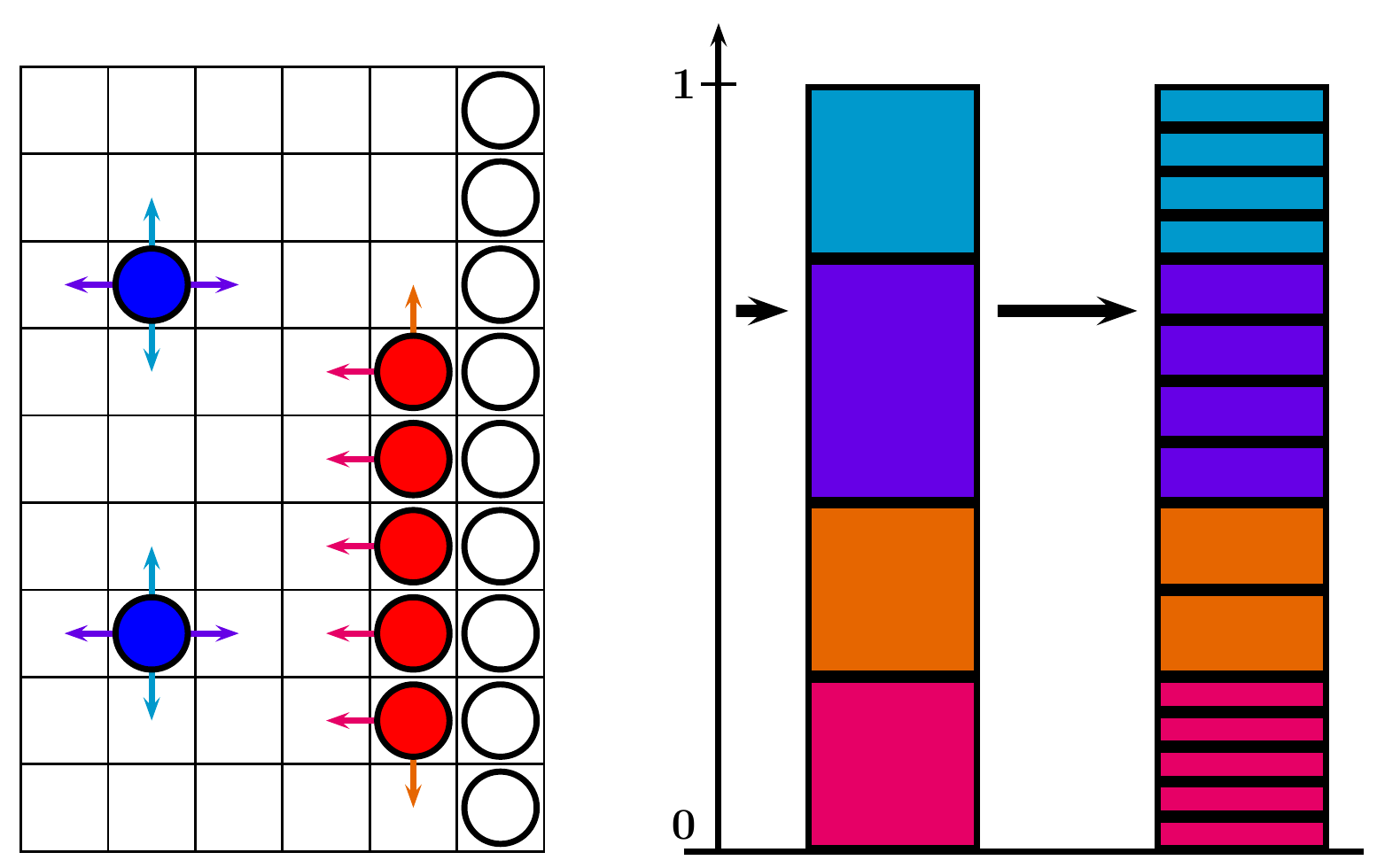} 
   \caption{Principle of process-type list algorithm. There are certain types of processes, indicated by colors in the figure: diffusion on the substrate, diffusion along a step, detachment from a step, \ldots (left scheme).
   Each type is described by its specific rate constant, but processes of the same type have the same rate constant. Hence, the list of all processes can be built up as a nested sum, first summing over processes of a given type, then over the various types. The selection of a process by a random number generator (right scheme) is realised in two steps, as indicated by the thick horizontal arrows, where the second one selects among equal probabilities.}
   \label{fig:process-types}
\end{figure}

The distinction between MC and kMC simulations is best understood by considering the following points:  In kMC, the possible configurations of the system, i.e. the micro-states contributing to the macro-state of a statistical ensemble, need to be enumerable, in order to be able to build up a list of process types, as in Table~\ref{tab:BKL}. In a MC simulation, on the other hand, there is no limit on the number of micro-states -- they even need not be known to start the simulation. 
For this reason, the MC algorithm can be applied to problems with a huge configuration space, e.g. to protein folding, where a kMC simulation would not be feasible.  
In advantage over MC, a kMC simulation allows us to assign the meaning of a physical time to the simulation steps. Of course, in order to make use of this advantage, we need to provide as input the rates of all relevant individual processes. 
Obtaining information about all these rates is a difficult task; this is why kMC simulations are less common than MC simulations. 
The best way for getting values for the individual rates is by performing molecular dynamics simulations,
possibly with first-principles electronic structure methods such as DFT. This will be explained in more detail in Section~\ref{sec:accelerated}.

Finally, we note that a kMC simulation provides a particular solution of the master equation in a stochastic sense; by averaging over many kMC runs we obtain probabilities (for the system being in a specific micro-state) that evolve in time according to Eq.~(\ref{master_eq}). 

\subsection{Algorithms for kMC}
In the kMC algorithm outlined above, the process list is ordered according to the process type; therefore I refer to it as the {\bf process-type list} algorithm.
In the practical implementation of kMC algorithms, there are two main concerns that affect the computational efficiency: First, the selection of a suitable algorithm depends on the question {\em how many} process types are typically active at each moment in time. The second concern is to  find an efficient scheme of representing and updating the data. 
For an efficient simulation, it is essential to realise  that the updating of the process list, step 4 of the algorithm described in the previous Section, only modifies those entries that have changed due to the preceding simulation step. A complete re-build of the list after each simulation step would be too time-consuming.  As the interactions are short-ranged, a local representation of the partial rates associated with lattice sites is most efficient.
On the other hand, the process-type list groups together processes having the same local environment, 
disregarding where the representatives of this class of processes (the spins or atoms) are actually located on the lattice. 
Hence, updating the process list requires replacement of various entries that originate from a small spatial region, but are scattered throughout the process list. 
To handle this task, a subtle system of referencing the entries is required in the code. This is best realised in a computer language  such as $C$ by means of pointers.
An example of a kMC code treating the SOS model is available from the author upon request.

The two-step selection of the next event, as illustrated in Fig.~\ref{fig:process-types}, 
makes the process-type list advantageous for simulation with a moderate number (say, $N < 100$) of process types. This situation is encountered in many simulations of epitaxial crystal growth using an SOS model\cite{Gilm76}, but the process-list algorithm also works well for more refined models of crystal growth\cite{KrSc:02,KrPe:02}.
In the first selection step, we need to compare the random number $\rho_1$ to at most $N$ partial sums, namely the expressions $\sum_i^k \, n_i w_i$ for $k=1, \ldots N$. The second selection step chooses among equally probable alternatives, and requires no further comparing of numbers. Thus, the total number of numerical comparisons needed for the selection is at most $N$, assuring that this selection scheme is computationally efficient.

In some applications, the kMC algorithm needs to cope with a vast number of different process types. 
For example, such a situation is encountered in epitaxy when the interaction is fairly long-ranged\cite{FiSc00}, or when rates depend on a continuous variable, such as the local strain in an elastically deformed lattice. Having to choose among a huge number of process types makes the selection based on the process-type list inefficient. Instead, one may prefer to work directly with a local data representation, and to do the selection of a process in real space. One may construct a suitable multi-step selection scheme by  grouping the processes in real space, as suggested by Maksym \cite{Maks88}. Then, one will first draw a random number $\rho_1$ to select a region in space, then use a second random number $\rho_2$ to select a particular processes that may take place in this region. Obviously, such a selection scheme is independent of the number of process types, and hence can work efficiently even if a huge number of process types is accessible. Moreover, it can be generalized further: It is  always possible to select one event out of $N=2^k$ possibilities  by making $k$ alternative decisions. 
This comes with the additional effort of having to draw $k$ random numbers $\rho_i, i=1,\ldots k$, but has the advantage that one needs to compare to $k= \log_2 N$ partial sums only. The most efficient way of doing the selection is to arrange the partial sums of individual rates on a  {\bf binary tree}. This allows for a fast hierarchical update of the partial sums associated with each branch point of the tree after a process has been executed. 

Finally, I'd like to introduce a third possible algorithm for kMC simulations that abandons the idea of the $N$-fold way. Instead, it emphasizes the aspect that each individual event, in as far as it is independent from the others, occurs after a random waiting time according to a Poissonian distribution. I refer to that algorithm as the  {\bf time-ordered list} algorithm, but frequently it is also called the 'first reaction' method\cite{LuSe98,LeHo01}.  It proceeds as follows:
\begin{enumerate}
\item At time $t$, assign a prospective execution time $t+t_i$ to each individual event, drawing the random waiting time $t_i$ from a Poissonian distribution; 
\item sort all processes according to prospective execution time (This requires only $\log_2 N$ comparisons, if done in a 'binary tree'); 
\item always select the {\em first} process of the time-ordered list and execute it; 
\item advance the clock to the execution time, and update the list according to the new configuration.
\end{enumerate}
This algorithm requires the $t_i$ to be Poissonian random numbers, i.e. to be distributed between $0$ and $\infty$ according to an exponentially decaying distribution function. Hence it may be advisable to use a specially designed random number generator that yields such a distribution.
The time-ordered-list algorithm differs from the two others in the fact that the selection step is deterministic, as always the top entry is selected. Yet, its results are completely equivalent to the two other algorithms, provided the common assumption of Poissonian processes holds: In a Poissonian processes, the waiting times are distributed exponentially.\cite{FiWe91} In the time-ordered list algorithm, this is warranted explicitly for each event by assigning its time of execution in advance. In the other algorithms, the clock, i.e., the 'global' time for all events, advances according to a Poissonian process. The individual events are picked at random from a list; however, it is known from probability theory that drawing a low-probability event from a long list results in a Poissonian distribution of the time until this event gets selected. Hence, not only the global time variable, but also the waiting time for an 
individual event follows a Poissonian distribution, as it should be.

The time-ordered list algorithm appears to be the most general and straightforward of the three algorithms discussed here. But again, careful coding is required: As for the process-type list, updating the time-ordered 
list requires deletion or insertion of entries scattered all over the list. 
Suggestions how this can be achieved, together with a useful discussion of algorithmic efficiency and  some more variants of kMC algorithms can be found in Ref.~ \cite{LuSe98}.

In principle, kMC is an inherently serial algorithm, since in one cycle of the loop only one process can be executed, no matter how large the simulation area is. Nonetheless, there have been a number of attempts to design {\bf parallel kMC algorithms}, with mixed success. All these parallel versions are based on a partitioning, in one way or another, of the total simulation area among parallel processors. However, the existence of a global 'clock' in the kMC algorithm would prevent the parallel processors from working independently. In practice, most parallel kMC algorithms let each processor run independently for some  time interval small on the scale of the whole simulation, but still long enough to comprise of a large number of events.  After each time interval the processors are synchronised and exchange data about the actual configurations of their neighbours. 
Typically, this communication among processors must be done very frequently during program execution. Hence the parallel efficiency strongly depends on latency and bandwidth of the communication network.
There are a number of problems to overcome in parallel kMC: Like in any parallel simulation of discrete events, the 'time horizon' of the processors may proceed quite inhomogeneously, and processors with little work to do may wait a long time until other, more busy processors have catched up. Even a bigger problem may arise from events near the boundary of processors: Such events may turn out to be impossible after the synchronisation has been done, because the neighbour processor may have modified the boundary region prior to the execution of the event in question. Knowing the actual state of the neighbouring processor, the event should have occurred with a different rate, or maybe not at all. In this case, a 'roll-back' is required, i.e., the simulation must be set back to the last valid event before the conflicting boundary event occurred, and the later simulation steps must be discarded. While such roll-backs are manageable in principle, they may lead to a dramatic decrease in the efficiency of a parallel  kMC algorithm. Yet, one may hope that the problems can be kept under control by choosing a suitable synchronisation interval. This is essentially the idea behind the 
'optimistic' synchronous relaxation algorithm \cite{ShAm05a,MeFi07}.
Several variants have been suggested that sacrifice less efficiency, but pay the price of a somewhat sloppy treatment of the basic simulation rules. 
In the semi-rigorous synchronous sublattice algoritm \cite{ShAm05b}, the first, coarse partitioning of the simulation area is further divided into sublattices, e.g. like the black and white fields on the checkerboard. Then, in each time interval between synchronisations, events are alternatingly simulated {\em only} within one of the sublattices ('black or 'white'). This introduces an arbitrary rule additionally restricting the possible processes, and thus may compromise the validity of the results obtained, but it allows one to minimise or even completely eliminate conflicting boundary events.  
Consequently, 'roll backs' that are detrimental to the parallel efficiency can be reduced or avoided. 
However, even when playing such tricks, the scalability of parallel kMC simulations for typical tasks is practically limited to four or eight parallel processors with the currently available parallel algorithms.\cite{NaSh09}

\section{From molecular dynamics to kMC: the bottom-up approach}
So far, we have been considering model systems. In order to make the formalism developed so far useful for chemistry or materials science, we need to describe how the relevant processes and their rate constants can be determined in a sensible way for a certain system or material. This implies bridging between the level of a molecular dynamics description, where the system is described by the positions and momenta of all particles, and the level of symbolic dynamics characteristic of kMC. For a completely general case, this may be a daunting task. For the special case of surface diffusion and epitaxial growth, it is typically a complex, but manageable problem.
On the atomistic level, the motion of an adatom on a substrate is governed by the potential-energy surface (PES),
which is the potential energy experienced by the diffusing adatom 
\begin{equation}
E^{\rm PES}(X_{\rm ad}, Y_{\rm ad}) = \min_{Z_{\rm ad},\{{\bf R}_I\}}
U (X_{\rm ad}, Y_{\rm ad}, Z_{\rm ad},\{{\bf R}_I\}) \, .
\label{PES}
\end{equation} 
Here $U(X_{\rm ad}, Y_{\rm ad}, Z_{\rm ad},\{{\bf
  R}_I\})$ is the potential energy of the atomic configuration specified by the coordinates  $(X_{\rm ad}, Y_{\rm  ad}, Z_{\rm ad},\{{\bf R}_I\})$.  
According to Eq.~(\ref{PES}), the PES
is the minimum of the potential energy with respect to both the adsorption height, denoted by  $Z_{\rm ad}$, and all coordinates of the substrate atoms, denoted by $\{{\bf R}_I\}$. 
The potential energy $U$ can in principle be calculated from any theory of the underlying microscopic physics. 
Presently, calculations of the electronic structure using density functional theory (DFT) are the most practical means of obtaining an accurate PES.   Within DFT, the energy $U$ in Eq.~(\ref{PES}) is referred to as the total energy (of the combined system of electrons and nuclei). 
The above expression is valid at zero temperature. At realistic temperatures, the free energy should be considered instead of $U$.
If we assume for the moment that the vibrational contributions to the free energy do not change the topology of the PES significantly, the minima of the PES represent the stable and metastable sites of the adatom.  

Next, we need to distinguish between crystalline solids on the one hand, and amorphous solids or liquids on the other hand. For a crystalline substrate, one will frequently (but not always) encounter the situation that the minima of the PES can be mapped in some way onto (possibly a subset of) lattice sites. 
The lattice sites may fall into several different classes, but it is crucial that all lattice sites belonging to one class are always connected in the same way to neighbouring sites. 
Then the dynamics of the system can be considered as a sequence of discrete transitions, starting and ending at lattice sites (lattice approximation). The sites belonging to one class all have the same number of connections, and each connection, i.e. each possible transition, is associated with a rate constant. 
Methods for amorphous materials going beyond this framework will be discussed later in Section~\ref{sec:accelerated}.

\begin{figure}[htbp] 
   \centering
   \includegraphics[width=9cm]{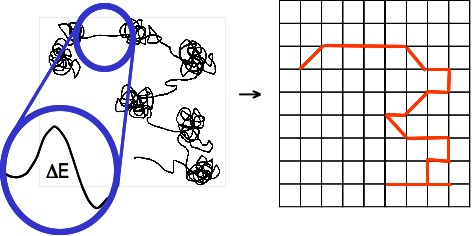} 
   \caption{Mapping of the diffusive Brownian motion of a particle on a substrate onto hopping between lattice sites. The particle's trajectory spends most of its time near the minima. In the blow-up of the small piece of the trajectory that crosses a saddle point between two minima, the energy profile along the reaction path is shown. Along the path, the saddle point appears as a maximum of the energy associated with the energy barrier $\Delta E$ that must be overcome by the hopping particle.}
   \label{fig:hopping}
\end{figure}

In surface diffusion, a widely used approximation for calculating the rate constants for the transitions between lattice sites is the so-called Transition State Theory (TST). As this is the 'workhorse' of the field, we will describe it first.
The more refined techniques presented later can be divided into two classes: techniques allowing for additional complexity but building on TST for the individual rates, and attempts to go beyond TST in the evaluation of rate constants.

\subsection{Determining rate constants from microscopic physics}
In order to go from a microscopic description (typical of a molecular dynamics simulation) to a meso- or macroscopic description by kinetic  theories, we start by dividing the phase space of the system into a 'more important' and a 'less important' part. In the important part, we'll persist to follow the motion of individual degrees of freedom. One such degree of freedom is the so-called 'reaction coordinate' that connects the initial state $x$ with a particular final state $x'$ (both minima of the PES)  we are interested in. The reaction coordinate may be a single atomic coordinate, a linear combination of atomic degrees of freedom, or, most generally, even a curved path in configuration space. The degrees of freedom in the 'less important' part of the system are no more considered individually, but lumped together in a 'heat bath', a thermal ensemble characterised by a temperature. 
In surface diffusion, the mentioned division of the system into 'more important' and 'less important' parts could (but need not) co-incide with the above distinction between the coordinates of the adsorbate atom,   $(X_{\rm ad}, Y_{\rm ad}, Z_{\rm ad})$, and the substrate atoms,  $\{{\bf R}_I\}$. 
Here, as in most cases, the distinction between the two parts is not unique; there is some arbitrariness, but retaining a sufficiently large part whose dynamics is treated explicitly should yield results that are independent of the exact way the division is made.
Of course,  the two parts of the system are still coupled: The motion along the reaction path may dissipate energy to the heat bath, an effect that is usually described by a friction constant $\lambda$. Likewise, thermal fluctuations in the heat bath give rise to a fluctuating force acting on the reaction coordinate.
\footnote{The friction force and the fluctuations of the random thermal force are interrelated, as required by the fluctuation-dissipation theorem.} 

Now we want to calculate the rate for the system to pass from the initial to the final state, at a given temperature of the heat bath. For the case of interest, the two states are separated by an energy barrier (or, at least, by a barrier in the free energy). For this reason, the average waiting time for the transition is much longer than typical microscopic time scales, e.g. the period of vibrational motion of a particle in a minimum of the potential. In other words, the transition is an infrequent event; and trying to observe it in a molecular dynamics simulation that treats the whole system would be extremely cumbersome. Therefore, we turn to rate theory to treat such rare events. 
Within the setting outlined so far, there is still room for markedly different behaviour of different systems, depending on the coupling between the system and the heat bath, expressed by the friction constant $\lambda$, being 'strong' or 'weak'. For a general discussion, the reader is referred to a review article\cite{Hanggi:90}. In the simplest case, if the value of the friction constant is within some upper and lower bounds, one can show that the result for the rate is independent of the value of $\lambda$. This is the regime where Transition State Theory is valid\cite{Glasstone:41,Vineyard:57}. If the condition is met, one can
derive  a form of the rate law  
\begin{equation}
w_{i}= \frac{k_{\rm B}T}{h} \exp(-\Delta F_{i}/(k_{\rm B}T)) \, ,
\label{ratdef}
\end{equation} 
for the rate $w_i$ of a molecular process $i$. Here, $i$ is a shorthand notation for the pair of states $x, \, x'$, i.e., $w_i \equiv w(x'|x)$.
In this expression, $\Delta F_{i}$ is the difference in the free
energy between the maximum (saddle point) and the minimum (initial geometry)
of the potential-energy surface along the reaction path of the
process $i$. $T$ is the
temperature, $k_{\rm B}$ the Boltzmann constant, and $h$ the Planck constant.
Somewhat oversimplifying, one can understand the expression for the rate as consisting of two factors: The first factor describes the inverse of the time it takes for a particle with thermal velocity to traverse the barrier region. The second factor accounts for the (small) probability that a sufficient amount of energy to overcome the barrier is present in the reaction coordinate, i.e., various portions of energy usually scattered among the many degrees of freedom of the heat bath happen by chance to  be collected for a short moment in the motion along the reaction coordinate.  The probability for this (rather unlikely) distribution of the energy can be described by the Boltzmann factor in Eq.~(\ref{ratdef}). The assumptions for the applicability of TST imply that the free energy barrier $\Delta F_i$ should be considerably higher than $k_{\rm B}T$. Speaking in a sloppy way, one could say that for such high barriers 'gathering the energy to overcome the barrier' and 'crossing the barrier' are two independent things, reflected in the two factors in the TST expression of the rate. 

The free energy of activation $\Delta F_{i}$ needed by the system to move from the initial position to the saddle point may be expressed in two ways: Using the fact that the free energy is the thermodynamic potential of the canonical ensemble, $\Delta F_{i}$ may be expressed by the ratio of partition functions,
\be
\Delta F_{i}= k_{\rm B} \log \left( \frac{Z_i^{(0)}}{Z_i^{\rm TS}} \right) \, .
\label{eq:ratio_part_functions}
\ee
where $Z_i^{(0)}$ is the partition function for the $m$ 'important' degrees of freedom of the system in its initial state, and $Z_i^{\rm TS}$ is the partition function for a system with $m-1$ degrees of freedom located at the transition state (saddle point). This partition function must be evaluated with the constraint that only motion in the hyperplane perpendicular to the reaction coordinate are permitted; hence the number of degrees of freedom is reduced by one. 

Alternatively, one may use the decomposition 
\begin{equation}
\Delta F_{i} = \Delta E_{i} - T \Delta S^{\rm vib}_{i} \, .
\label{deltaf}
\end{equation} 
Here $\Delta E_{i}$ is the difference of the internal energy
(the (static) total energy and the vibrational energy) of the 
system at the
saddle point and at the minimum, and $\Delta S^{\rm vib}_{i}$ is the
analogous difference in the vibrational entropy. The rate of the process $i$
can be cast as follows, 
\begin{equation}
w_{i} = w^{(0)}_{i} \exp(-\Delta E_{i}/k_{\rm B}T) \, ,
\label{retdef1}
\end{equation} 
where $w^{(0)}_{i} = (k_{\rm B}T/{h}) \exp (\Delta S^{\rm
  vib}_{i}/k_{\rm B})$ is called the attempt frequency. 

The two basic quantities in Eq. (\ref{retdef1}), 
$w^{(0)}_{i}$ and $\Delta E_{i}$,
can both be obtained from DFT calculations. If we restrict ourselves to 
single-particle diffusion and neglect the contribution of thermal vibrational energy, $\Delta E_{i}$ can be 
read off directly from the PES.  
To obtain the value of the attempt frequency, one may perform molecular dynamics simulations of the canonical ensemble, sampling the partition functions $Z^{(0)}_i$ and $Z^{\rm TS}_i$. 
For a computationally simpler, but less accurate approach, one may expand the PES in a quadratic form around the minimum and the saddle point. In this approximation, the partition functions in Eq.~(\ref{eq:ratio_part_functions}), which then equal those of harmonic oscillators, may be evaluated analytically, and one arrives at the frequently used expression
\be
w^{(0)}_i = \frac{ \prod_{k=1}^n \omega^{(0)}_{k,i}}{ \prod_{k=1}^{n-1} \omega^{\rm TS}_{k,i}} \, .
\ee 
Here $\omega^{(0)}_{k,i}$ and $\omega^{\rm TS}_{k,i}$ are the frequencies of the normal modes at the initial minimum and at the transition state of process $i$, respectively. Note that the attempt frequency, within the harmonic approximation, is independent of temperature.

Finally, we will briefly comment on the validity of TST for processes in epitaxial growth. 
For surface diffusion of single adatoms, it has been shown for the case of Cu/Cu(111) that TST with thermodynamic sampling of the partition functions gives good results (i.e. in agreement with molecular dynamics simulations) for the temperature regime of interest in epitaxial growth experiments. The harmonic approximation is less satisfactory, but still yields the correct order of magnitude of the surface hopping rate\cite{Boisvert:98a}. For systems with low energy barriers ($< 3 k_{\rm B}T$), or for collective diffusion processes, it is generally difficult to judge the validity of TST. 
In the latter case, even locating all saddle points that can be reached from a given initial state is a challenge. For this task, algorithms that allow for locating saddle points without prior knowledge of the final state have been developed. The 'dimer' method\cite{HeJo99} is an example for such a method. It is well suited for being used together with DFT calculations, as it requires only first (spatial) derivatives of the PES, and is robust against numerical errors in the forces.

\subsection{Accelerated molecular dynamics \label{sec:accelerated} }
In this Section, I'll briefly introduce methods that are suitable if the lattice approximation cannot be made, or if one needs to go beyond transition state theory. These methods employ some refinement of molecular dynamics that allows one to speed up the simulations, such that so-called 'rare' events can be observed during the run-time of a simulation. In this context, 'rare' event means an event whose rate is much smaller than the frequencies of vibrational modes. Keeping the TST estimate of rate constants in mind, any process that requires to overcome a barrier of several $k_{\rm B}T$ is a 'rare' event. Still it could take place millions of times on an experimental time scale, say, within one second. Therefore 'rare' events could be very relevant for example for simulations of epitaxial growth. Ref.~\cite{VoMo02} provides a more detailed overview of this field.

Obviously, running simulations in parallel is one possible way to access longer time scales. In the {\bf parallel replica method}\cite{Vote98}, one initiates several parallel simulations of the canonical ensemble starting in the same initial minimum of the PES, and observes if the system makes a transition to any other minimum. Each replica runs independently and evolves  differently due to different fluctuating forces. From the abundances of various transitions observed during the parallel run, one can estimate the  rate constants of these transitions, and give upper bounds for the rates of possible other transitions that did not occur during the finite runtime. 
The method is computationally very heavy, but has the advantage of being unbiased towards any (possibly wrong) expectations how the relevant processes may look like.

Another suggestion to speed up molecular dynamics simulations goes under the term 
{\bf hyperdynamics}\cite{Vote97}. The 'rare event problem' is overcome by adding an artificial potential to the PES that retains the barrier region(s) but modifies the minima so as to make them shallower. The presence of such a 
'boost potential'  will allow the particle to escape from the minimum more quickly, and hence the processes of interest (transitions to other minima) can be observed within a shorter MD run. 
The method can be justified rigorously for simulations where one is interested in thermodynamic equilibrium properties (e.g., partition function): The effect of the boost potential can be corrected for by introducing a time-dependent weighting factor in the sampling of time averages. It has been suggested to extend this approach beyond thermal equilibrium to kinetical simulations:
While the trajectory passes the barrier region unaffected by the boost potential, the simulation time corresponds directly to physical time. 
While the particle stays near a minimum of the PES, and thus under the influence of the boost potential, its effect must be corrected by making the physical time to advance faster than the simulation time. Ways to construct the boost potential in such a way that the method yields unchanged thermal averages of observables have been devised\cite{Vote97}. However, it has been argued that the speed-up of a simulation of epitaxy achievable with such a global boost potential is only modest if the system, as usually the case, consists of many particles\cite{HeJo01}.
This restriction can be overcome by using a local boost potential\cite{WaPa01,MiFi04} rather than a global one. In this case it is assumed that the transitions to be 'boosted' are essentially single-particle hops.  This, of course, curtails one strength of accelerated MD methods, namely being unbiased towards the (possibly wrong) expectations of the users what processes should be the important ones.  Also, it is important to note that 
the procedure for undoing the effect of the boost potential relies on assumptions of the same type as TST. Therefore hyperdynamics cannot be used to calculate rate constants more accurately than TST.

To be able to observe more transitions and thus obtain better statistics within the (precious) runtime of an MD simulation, people have come up with the simple idea of increasing the simulation temperature. This approach is particularly attractive if one wants to simulate a physical situation where the temperature is low, e.g. low-temperature epitaxial growth of metals. By running at an artificially raised temperature (For solids, the  melting temperature is an upper bound), a speed-up by several orders of magnitude may be achieved.  Of course, the physics at high and low temperatures is different, thus invalidating a direct interpretation of the results obtained in this way. However, combining the idea of increased temperature MD with the principles used in kMC simulations provides us with a powerful tool. It comes under the name of 
{\bf temperature-accelerated MD}\cite{SoVo00}, abbreviated as TAD: First, a bunch of MD simulations is  performed, starting from the same initial state (as in the parallel replica method), at a temperature $T_{\rm high}$ higher than the physical temperature. The transitions observed during these runs are used for estimating their individual rates and for building  up a process list. At this stage, TST in the harmonic approximation is used to 'downscale' the rate constants from their high-temperature value obtained from the MD simulation to their actual value at the lower physical temperature $T_{\rm low}$. If a process is associated with an energy barrier $\Delta E_{i}$, its rate constant should be scaled with a factor $\exp(\Delta E_{i} (T_{\rm high}^{-1} - T_{\rm low}^{-1})/k_{\rm B} )$. 
Having sampled many MD trajectories, it is also possible to provide an upper bound for the probability that some possibly relevant transition has not yet occurred in the available set of trajectories. 
In other words, in TAD simulations some kind of 'safe-guarding' can be applied not to overlook possibly important transitions. After sufficiently many trajectories have been run, a pre-defined confidence level is reached that the transitions observed so far are representative for the physical behaviour of the system in the given initial state, and can be used as a process list. 
Next, a kMC step is performed by selecting randomly one of the processes with probability proportional to the  (scaled) rates in the process list. Then the selected process is executed, and the system's configuration changes to a new minimum. The loop is closed by going back to the first step and performing MD simulations for the system starting from this new minimum, and attempting new transitions from there. 

Some more comments may be helpful. 
First, we note that the scaling factor used for downscaling the rates is different for different processes. 
Thus, the method accounts for the fact that the relative importance of high-barrier processes and low-barrier processes must be different at high and low temperatures, respectively. This requirement of a physically meaningful kinetical simulation would be violated by just naively running MD at a higher temperature without applying any corrections, but TAD passes this important test. 
Secondly, TAD may even provide us with information that goes beyond TST. For instance, if collective diffusion processes play a role, the relative abundance with which they were encountered in the MD runs gives us a direct estimate of the associated attempt frequency, without having to invoke the (sometimes questionable) approximations of TST. 
\footnote{The assumption that TST can be used for downscaling is a milder one than assuming the applicability of TST for the attempt frequency as such.} 
Third, 
one can gain in computational efficiency by using the same ideas as in kMC: 
The process list need not be build from scratch each time, but only those entries that changed since the last step need to be updated. 

Using this strategy, TAD has been applied to simulations of epitaxy\cite{MoSo01}. In this context, it should be noted that the need for starting MD simulations in each simulation step can be reduced further: As mentioned above, kMC is based on the idea that the local environment of a particle, and thus the processes accessible for this particle, can be classified. Once the TAD simulations have established the rates for all processes of a certain environmental class (e.g. terrace diffusion), these rates can be re-used for all particles in this class (e.g., all single adatoms on a terrace). This reduces the computational workload considerably. 

Finally, we mention that TAD has recently been 
combined with parallel kMC simulations using the semi-rigorous synchronous sublattice algorithm\cite{ShAm07}.

\section{Tackling with complexity\label{sec:beyond}}
In the early literature of the field, kMC simulations are typically considered as a tool to rationalize experimental findings. In this approach, one works with models that are as simple as possible, i.e., comprise as few process types as possible, while still allowing for reproducing the experimental data. The rates of these processes are then often treated as parameters whose values are adjusted to fit the data. The aim is to find a description of the experimental observations with a minimum number of parameters.

More recently, the focus has shifted to 
kMC simulations being perceived as a scale-bridging simulation technique that enables researchers to describe a specific material or materials treatment as accurately as desired. The goal is to perform kMC simulations where 
{\em all relevant} microscopic processes are considered, aiming at simulations with predictive power. 
This could mean that 
all distinct processes derived from a given Hamiltonian, e.g. an SOS model, should be included. However, for predictive simulations, a model Hamiltonian is often an already too narrow basis. The ultimate benchmark are (possibly accelerated) MD simulations that allow for an unbiased assessment  which processes are relevant for a specific material, and then to match the kMC simulations to these findings.

\begin{figure}[htbp] 
   \centering
   \includegraphics[width=4cm]{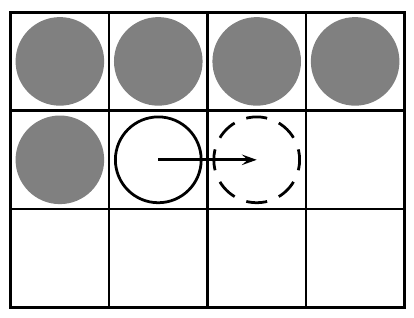} 
   \caption{Illustration of the local environments of a hopping particle (white circle, in its initial (full line) and final (dashed line) state) in a model with nearest-neighbor interactions. The classification may depend on the occupation of the sites adjacent to the initial or the final state. A particular occupation is indicated by the grey circles. Since each of the ten relevant sites may be either occupied or empty, there can be up to $2^{10}$ environment classes.}
   \label{fig:voter}
\end{figure}

As explained in the preceding Section, the efficiency of kMC simulations rests on a classification scheme for the local environments a particle encounters during the course of a simulation:  Wherever and whenever the particle is in the 'same' environment, the same process types and rate constants will be re-used over and over again.
However, the number of different process types to be considered may be rather big. 
For example, even for the simple SOS model, the complete process list could have up to  $~2^{10}$ entries~\cite{Vote86} (as explained below). 
This raises the issue of complexity: Apart from approximations  for calculating rate constants (such as TST), a kMC simulation may be more or less realistic depending on 
whether the classification of local environments and processes is very fine-grained, or whether a more coarse classification scheme is used. 

On one end of the complexity scale, we find kMC simulations that do not rely on a pre-defined process list. Instead, the accessible processes are re-determined after each step, i.e., the process list is generated 'on the fly' while the simulation proceeds. This can be done for instance by temperature-accelerated molecular dynamics (see preceding Section). If one is willing to accept TST as a valid approximation for the calculation of rate constants, molecular dynamics is not necessarily required; instead, it is computationally more efficient to perform a {\bf saddle point search}, using a modified dynamics for climbing 'up-hill' from a local minimum. An example for a saddle point search algorithm that uses such 'up-hill climbing' is the 'dimer' method\cite{HeJo01}. The overall design of the kMC algorithm employing the 'search on the fly' is similar to TAD: Starting from an initial state, one initiates a bunch of saddle point searches. For each saddle point encountered, TST is used to calculate the associated rate constant. If repeated searches find the known saddle points again and again with a similar relative frequency, one can be confident that the transitions found so far make up the complete process list for this particular initial state. Next, a kMC step is carried out, leading to a new configuration; the saddle point search is continued from there, etc. 

List-free kMC has been used to study metal epitaxy. For aluminum, for instance, these studies have revealed the importance of collective motion of groups of atoms for the surface mass transport\cite{HeJo03}.
We note that the lattice approximation is not essential for this approach.
Thus, it could even be applied to investigate kinetics in amorphous systems. 
While the saddle point search is considerably faster than MD, the method is, however, still orders of magnitude more expensive than list-directed kMC, in particular if used in conjunction with DFT to obtain the potential energy surface and the forces that enter the saddle point search.

The above method becomes computationally more affordable if the lattice approximation is imposed. The constraint that particles must sit on lattice sites reduces the possibility of collective motions, and thus invoking the lattice approximation  makes the method less general. On the other hand, using a lattice makes it easier to re-use rate constants calculated previously for processes taking place in the 'same' local environment. 
A variant of kMC called 'self-learning'\cite{TrKa05,KaAl06,KaTr09} also belongs into this context: Here, one starts with a pre-defined process list, but the algorithm is equipped with the ability to add new processes to this list if it encounters during the simulation a local environment for which no processes have been defined so far. In this case, additional saddle point searches have to be performed in order to obtain the rate constants to be added to the list.

At a lower level of complexity, we find kMC simulations that employ, in addition to the lattice approximation, a finite-range model for the interactions between particles. For the local minima of the PES, this implies that the depths of the minima can be described by a lattice Hamiltonian. 
For each minimum, there is an on-site energy term. If adjacent sites are occupied, the energy will be modified by pair interactions, triple interactions, etc. In materials science, this way of representing an observable in terms of the local environments of the atoms is called cluster expansion method.

The usage of a lattice Hamiltonian or cluster expansion is in principle an attractive tool for tackling with the complexity in a kMC simulation of crystalline materials. However, 
for calculating rate constants, we need (in TST) the energy {\em differences} between the transition state and the initial minimum the particle is sitting in. This complicates the situation considerably. 
To discuss the issue, let's assume that the interactions between particles are limited to nearest neighbors. Then, both the initial state and the final state of the particle can be characterized completely by specifying which of their neighbors are occupied. On a 2D square lattice, a particle  moving from one site to a (free) neighboring site has a shell of ten adjacent sites that could be either occupied or free (see Fig.~\ref{fig:voter}). Thus, the move is completely specified (within the nearest-neighbor model) by one out of $2^{10}$ possible local environments\cite{Vote86}. \footnote{To be precise, the actual number is somewhat smaller due to symmetry considerations.}
One way to specify the input for a kMC simulation is to specify a rate constant for each of these $~2^{10}$ process types. This is in principle possible if an automated algorithm is used to determine the energy barrier and attempt frequency for each case. 
For practical purposes, one may specify only a selected subset of the $~2^{10}$ rate constants, and assume that the rest takes on one of these specified values. This is equivalent to assuming that, at least for some environments, the occupation of some of the ten sites doesn't matter.
This approach has been used by the author to describe the rate constants for epitaxy on a semiconductor surface, GaAs(001)\cite{KrSc:02}. A process list with about 30 entries was employed to describe the most relevant process types. 

Another way of tackling with the complexity is the assumption that 
$\Delta E$ does not depend on the ocupation of sites, but only on the {\em energies} of the initial and final minima. The technical advantage of this approach lies in the fact that the energies of the minima may be represented via a lattice Hamiltonian (or, equivalently, by the cluster expansion method). Thus, these energies can be retrieved easily from the cluster expansion. However, there is no rigorous foundation for such an assumption, and its application could  introduce  uncontrolled approximations. 
For a pair of initial and final states, $i=(\mathrm{ini}, \mathrm{fin})$, one could, for example, assume that $\Delta E = \Delta E_0 + {1\over 2}(E_{\mathrm{fin}} - E_{\mathrm{ini}}) $. This assumption has been employed for diffusion of Ag adatoms on the Ag(111) surface in the presence of interactions\cite{FiSc00}, and test calculations using DFT for selected configurations of adatoms have confirmed its validity. Note that 
the dependence on the sum of the initial and final state energy assures that the forward and backward rate fulfill detailed balance, Eq.~(\ref{eq:detailedB}), as required for a physically meaningful simulation. 

In a large part of the literature on kMC, an even simpler assumption is made, and the rate constants are assumed to depend on the energy of the initial state only. In other word, the transition states for {\em all} processes are assumed to be at the same absolute energy. This assumption facilitates the simulations, but clearly is not very realistic. At this point, we have reached the opposite end on the scale of complexity, where the goal is no longer a realistic modeling of materials, but a compact description of experimental trends. 

I would like to conclude this Section with a word of caution: 
In epitaxial growth,  fine details of the molecular processes may have drastic consequences on the results of the simulations. Often, the details that make a difference are beyond the description by a lattice Hamiltonian. One example is the mass transport between adjacent terraces by particles hopping across a surface step. In many metals, the energy barrier for this process is somewhat higher than the barrier for conventional hopping diffusion on the terraces. This so-called Schw{\"o}bel-Ehrlich effect is crucial for the smoothness or roughness of epitaxially grown films, but is not accounted for by the SOS model. Thus, the rate for hopping across steps needs to be added 'by hand' to the process list of the SOS model to obtain sensible simulation results. 
Another example concerns the shape of epitaxial islands on close-packed metal surfaces, for instance Al(111) and Pt(111).
Here, either triangular or hexagonal islands can be observed, depending on the temperature at which an experiment of epitaxial growth is carried out. A detailed analysis shows that the occurrence of triangular islands is governed by the process of corner diffusion: An atom sitting at the corner of a hexagonal island, having an island atom as its only neighbor, has different probabilities for hopping to either of the two island edges\cite{BoSt98,OvBo99}. For this reason, there is a tendency to fill up one particular edge of a hexagonal island, and the island gradually evoles to a triangular shape. Only at higher temperatures, the difference between the two rates becomes less pronounced, and the hexagonal equilibrium shape of the islands evolves. Only with the help of DFT calculations it has been possible to detect the difference of the energy barriers for the two processes of corner diffusion.  Simplified models based on neighbor counting, however,  cannot detect such subtle differences, in particular if only the initial state is taken into account. Therefore, kinetic Monte Carlo studies addressing morphological  evolution should always be preceded by careful investigations of the relevant microscopic processes using high-level methods such as DFT for calculating the potential energy profiles. 

\section{Summary}
With this tutorial I intended to  familiarise the readers with the various tools to carry out scale-bridging simulations. These tools range from accelerated molecular dynamics simulations that extend the idea of Car-Parrinello molecular dynamics to longer time scales, to abstract models such as the lattice-gas Hamiltonian. The scientist interested in applying one of these tools should decide whether she/he wants to trust her/his intuition and start from an educated guess of a suitable kinetic model, such as SOS or similar. 
Else, she/he may prefer to 'play it safe', i.e. avoid as much as possible the risk of overlooking rare, but possibly important events.  In the latter case, kMC simulations in combination with saddle point searches (that build up the rate list 'on the fly') are a good choice. However, this methods could be computationally too expensive if slow changes in a system very close to equilibrium should be studied, or if vastly different processes play a role  whose rates span several orders of magnitude. In this case, considerations of numerical efficiency may demand from the user to make a pre-selection of processes that will be important for the evolution of the system towards the non-equilibrium structures one is interested in.  Using the $N$-fold way kinetic Monte Carlo algorithm with a pre-defined list of process types  could be a viable solution for these requirements. In summary, Monte Carlo methods allow one to go  in either direction, to be as accurate as desired (by including sufficiently many many details in the simulation), or to find a description of nature that is as simple as possible.

\section*{Acknowledgments}
I'd like to acknowledge helpful discussions with Wolfgang Paul, with whom I had the pleasure of organizing a joint workshop on kinetic Monte Carlo methods. Matthias Timmer is thanked for reading the  manuscript and making suggestions for improvements.


\end{document}